\documentclass[fleqn,usenatbib]{mnras}

\usepackage{newtxtext,newtxmath}

\usepackage[T1]{fontenc}

\DeclareRobustCommand{\VAN}[3]{#2}
\let\VANthebibliography\thebibliography
\def\thebibliography{\DeclareRobustCommand{\VAN}[3]{##3}\VANthebibliography}

\usepackage{graphicx}	
\usepackage{amsmath}	
\usepackage{booktabs}	
\usepackage{multirow}	
\usepackage{stfloats}	

\makeatletter
\def\mnras@float#1[#2]{\@float{#1}[#2]\csname fstyle@#1\endcsname}
\def\mnras@dblfloat#1[#2]{\@dblfloat{#1}[#2]\centering\csname fstyle@#1\endcsname}
\def\figure{\def\baselinestretch{1}\let\@makecaption=\SFB@makefigurecaption
  \@ifnextchar[{\mnras@float{figure}}{\@float{figure}\fstyle@figure}}
\@namedef{figure*}{\def\baselinestretch{1}\st@rredfloattrue
  \let\@makecaption=\SFB@makefigurecaption
  \@ifnextchar[{\mnras@dblfloat{figure}}{\@dblfloat{figure}\centering\fstyle@figure}}
\def\table{\def\baselinestretch{1}\let\@makecaption=\SFB@maketablecaption
  \@ifnextchar[{\mnras@float{table}}{\@float{table}\fstyle@table}}
\@namedef{table*}{\def\baselinestretch{1}\st@rredfloattrue
  \let\@makecaption=\SFB@maketablecaption
  \@ifnextchar[{\mnras@dblfloat{table}}{\@dblfloat{table}\centering\fstyle@table}}
\makeatother

\title{Simulation-based inference of profile evolution for high-precision pulsar timing}

\author[S. C. Susarla et al.]{
Sai C. Susarla,$^{1}$\thanks{E-mail: s.c.susarla@soton.ac.uk (SCS)}
Gregory Ashton,$^{1}$
A. Parthasarathy,$^{2,3,4}$
K. De Man$^{5}$
and J. P. W. Verbiest$^{6}$
\\
$^{1}$School of Mathematical Sciences, University of Southampton, University Road, Southampton SO17 1BJ, UK\\
$^{2}$ASTRON, Netherlands Institute for Radio Astronomy, Oude Hoogeveensedijk 4, 7991 PD Dwingeloo, The Netherlands\\
$^{3}$Anton Pannekoek Institute for Astronomy, University of Amsterdam, Science Park 904, 1098 XH Amsterdam, The Netherlands\\
$^{4}$Max Planck Institut f\"{u}r Radioastronomie, Auf dem H\"{u}gel 69, 53121 Bonn, Germany\\
$^{5}$Kapteyn Astronomical Institute, University of Groningen, PO Box 800, NL-9700 AV Groningen, The Netherlands\\
$^{6}$Florida Space Institute, University of Central Florida, 12354 Research Parkway, Orlando, FL 32826, USA
}

\date{Accepted XXX. Received YYY; in original form ZZZ}

\pubyear{\the\year{}}

\begin{document}
\label{firstpage}
\pagerange{\pageref{firstpage}--\pageref{lastpage}}
\maketitle

\begin{abstract}
Pulsars are exceptional cosmic laboratories, the most stable of which rival atomic clocks. Nearly all of that precision rests on one observable, the average pulse profile unique to each pulsar and stable over decades. These profiles change shape systematically with observing frequency, reflecting the physics of the emitting pulsar beam, but if left unmodelled, it can bias the arrival-time measurements, undermining precision timing. Profile-domain timing has been proposed to address this effect, but current implementations use stochastic sampling, which is computationally prohibitive and has limited the application in large-scale analyses. Here, we present a simulation-based inference approach to profile-domain timing, using a normalising flow conditioned on a convolutional embedding of the frequency-phase portrait. From a single portrait, it returns joint posteriors on dispersion measures, arrival times, and the parameters governing profile evolution instantaneously, with uncertainties verified against known inputs following an initial  training run. The value of a likelihood-free approach is that the simulator becomes the interface to pulsar properties. Fundamental properties like non-Gaussian noise, pulsar self-noise, jitter, scattering, and profile change can be inferred by simulating them rather than by deriving a new likelihood for each, opening a promising pathway to include machine learning methods into pulsar timing workflows.
\end{abstract}




\section{Introduction}\label{sec:intro}

The pulse profile of a radio pulsar evolves with observing frequency. Lower frequencies are emitted at greater heights in the magnetosphere, where the open magnetic field lines have diverged further, so the emission is beamed more broadly and the observed pulse is wider. This radius-to-frequency mapping \citep{Cordes1978, LorimerandKramer2005} makes the intrinsic profile shape a function of frequency. In addition to that, propagation through the interstellar medium \citep{keith2013} contributes to the profile evolution by dispersion or scattering. If left unmodelled, this evolution biases the time-of-arrival (ToA) measurements on which high-precision timing depends \citep{verbiest2018}. As instrumentation has advanced, with wideband receivers spanning ever larger bandwidths, the profile evolution within a single observation becomes even harder to model \citep{Pennuccietal2014, Liu2014}. This is exacerbated at low-frequency telescopes like LOFAR \citep{vanhaarlem2013} and, in future, the Square Kilometre Array (SKA) \citep{braun_ska,keane2018}. For the pulsar timing arrays (PTAs) now hunting the nano-hertz gravitational-wave background \citep{eptagwb,agazie2024}, unmodelled profile evolution is among the systematics that limit achievable timing precision \citep{Tiburzi2015,verbiest2018}.

\begin{figure*}
    \centering
    \includegraphics[width=\linewidth]{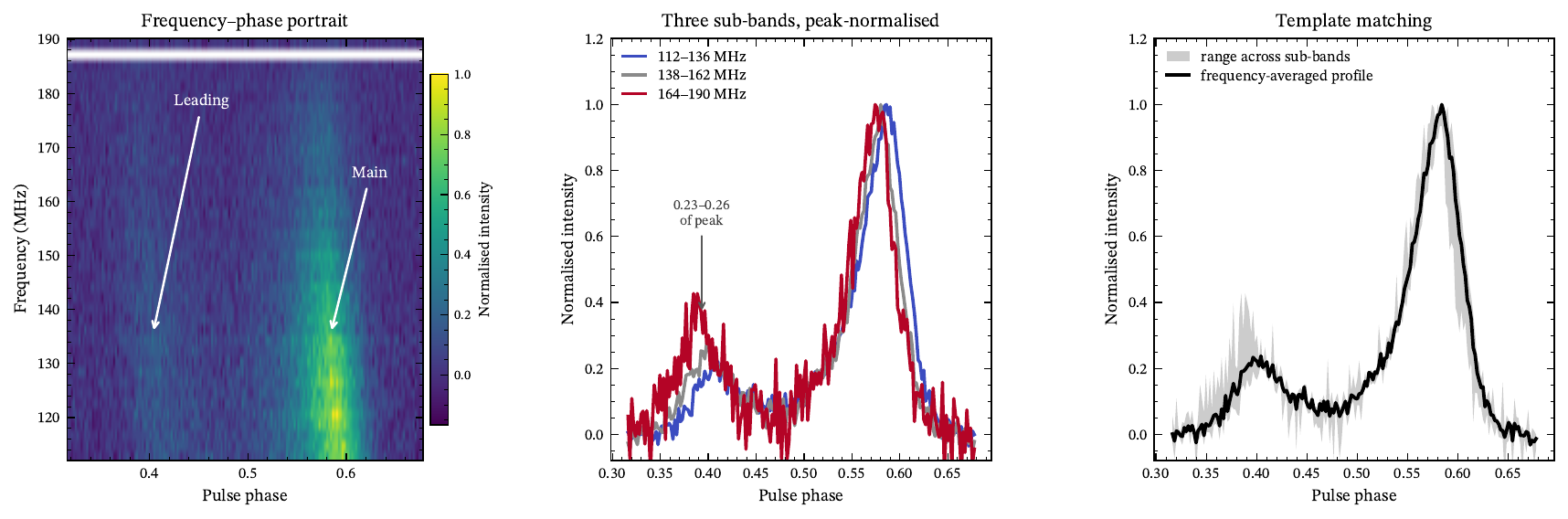}
    \caption{Frequency evolution of PSR~J2145$-$0750 observed using Irish-LOFAR telescope \citep{REALTA_2021}. The first panel shows the frequency--phase portrait and the two prominent pulses (the white band is a dead channel). The second panel shows the flux profiles for three different subbands. The third panel shows an example of the template used in the template-matching method in which a cross-correlation is used to estimate the ToA.}
    \label{fig:introfig}
\end{figure*}

The conventional approach to pulsar timing, known as \emph{template matching} \citep{taylor1992}, cross-correlates each observation against a fixed, frequency-collapsed template to produce a pulse ToA \citep{vanstraten2006}. It is fast and simple, but ignores the frequency-resolved structure of the data and is correspondingly vulnerable to profile variation \citep{kramer1998}. However, with the advent of wide-band observing systems (like those on LOFAR, MeerKAT \citep{Meerkat2020}, CHIME \citep{Chime2022}, and Parkes \citep{Parkes2020}), time-variable dispersion measures (DMs) have driven two independent changes to the template-matching approach. First, the advent of frequency-resolved templates that are built either by combining observations \citep{donnerfddm2019, donner2020} or by modelling the two-dimensional profile directly \citep{Pennuccietal2014, Pennucci2019}. The second concerns the ToAs themselves, where each channel may be timed independently (narrowband), or the whole observation matched to a frequency-resolved template in a two-dimensional fit determining ToA and DM simultaneously (wideband). The two choices are separable, and both are now in routine use across pulsar timing arrays \citep{agazie2024, paladi2024, wideband2025}. In every case, however, the template is built once, generally from the addition of all the epochs in the dataset, and is held fixed afterwards. These approaches cannot accommodate epoch-dependent effects such as time-variable scattering common at these frequencies, returns no inference on the profile evolution parameters themselves, and the uncertainty in the template does not propagate into the reported ToA and DM uncertainties.

To address the inadequacy of conventional timing methods, profile-domain timing was introduced, where we infer the timing parameters using shapelet decompositions in a fully Bayesian framework \citep{Lentati2015, lentati2017}, and can accommodate epoch-to-epoch profile variation. This comes at a cost, as its stochastic-sampling implementations are computationally intensive, and adoption has remained sparse relative to its potential. To this day, every operating PTA relies heavily on template matching \citep{verbiest2018}, with its known vulnerability to unmodelled profile changes. Fig~\ref{fig:introfig} shows the frequency evolution in PSR~J2145$-$0750, where we highlight the leading and main pulses. The middle panel shows the evolution, particularly in the leading component and the shift in phase due to the dispersion sweep. The third panel shows the template used in the traditional template-matching method, which doesn't account for the frequency evolution happening in the entire band.  

In this work, we apply simulation-based inference (SBI), a class of methods that learn to perform Bayesian inference from simulations alone, requiring no tractable likelihood \citep{Cranmer2020}, to the pulsar timing problem for the first time, making profile-domain timing fast enough for routine use at scale.
We use neural posterior estimation \citep{Papa2016,greenberg2019,tejero2020}, training a normalizing flow \citep{Durkan2019} on realistic simulations of a pulsar's two-dimensional frequency--phase portrait. A convolutional embedding network compresses each portrait, and the flow learns the structure of the profile evolution directly.
SBI has seen rapid uptake across astronomy, from parameter estimation in gravitational-wave (GW) astronomy \citep{Dax2021} to cosmological inference \citep{SpurioMancini:2021ppk} and studies of galaxy clustering \citep{SimBIG:2023ywd}.
Within pulsar timing specifically, it has so far been applied to adjacent problems, such as GW inference from PTA data \citep{ShihPTA,ChenPTAML} and population-synthesis-based parameter estimation of isolated pulsars \citep{SBIpulsarpopu}.
Here, we extend SBI to the pulsar timing problem itself for the first time, applying it to profile-domain timing of individual pulsars.

To test this method, we chose PSR~J2145$-$0750 as it is a well-known PTA millisecond pulsar. It has exceptional timing stability, with predictable frequency evolution in the LOFAR band and no significant profile change events reported. Next, we simulate the pulsar's frequency evolution across the LOFAR high band spanning the frequency range 112--190 MHz. This pulsar has two main on-pulse regions which can be depicted using four components. We introduce a set of parameters describing profile evolution in the frequency--phase domain. Each simulation is characterised by six parameters: two timing observables, the dispersion measure residual $\Delta$DM and the ToA, expressed as a phase offset; two spectral indices, $\gamma_\mathrm{lead}$ and $\gamma_\mathrm{main}$, setting how the amplitudes of the leading and main profile components scale with frequency; $A_\mathrm{ref}$, the amplitude of the leading component relative to the main component at $\nu_\mathrm{ref}$; and the signal-to-noise ratio, parametrised as $\log(\mathrm{S/N})$. A further set of parameters describing component widths, secondary amplitudes and per-channel instrumental gain is drawn from priors at each simulation and marginalised over rather than inferred, so that the posterior accounts for uncertainty in these quantities. The inference network is performed by a normalizing flow conditioned on a multi-branch convolutional neural network (CNN) with summary statistics derived from the portrait, which are used to resolve the degeneracies between various parameters.

Applied to observations of the millisecond pulsar PSR~J2145$-$0750 at Irish-LOFAR High Band Antenna~\citep{REALTA_2021,susarla_2025_dr}, the trained network reproduces the DMs of a conventional epoch-wise \citep{iraci2024} analysis to within their combined uncertainties, while additionally constraining the parameters that describe how the profile evolves across the band, quantities a frequency-collapsed template cannot access. This method constrains parameters of the profile evolution and spectral indices of amplitudes of individual components simultaneously. Because inference is amortised, a single training run can be used for any number of observations at negligible cost. Therefore, profile-domain timing using SBI becomes practical at the scale of current and forthcoming PTA datasets. The structure of the rest of the article is as follows: Sect.~\ref{sec:methods} shows the methodology of the forward simulator and the embedding architecture, Sect.~\ref{sec:valcal} shows the validation and calibration of the methodology on simulations, Sect.~\ref{sec:results} shows the application to the real data and consistency checks and Sect.~\ref{sec:conclusions} shows the discussion, conclusions and the future prospects.

\section{Methods}\label{sec:methods}
To carry out SBI, the most important part is the simulations: inference on the posteriors depend entirely on how accurately the simulations model the real observing setup. We built a forward simulator that mimics the profile evolution of PSR~J2145$-$0750. It generates synthetic frequency--phase portraits by constructing an analytic pulse model, applying interstellar medium (ISM) propagation effects, and adding per-pixel Gaussian noise, calibrated so that the frequency-averaged profile reaches a target S/N. The pulsar selection was motivated by the fact that it shows almost no reported profile change events and is well timed and observed by numerous telescopes worldwide. The forward simulator is specific to this pulsar at this frequency band. Each simulated frequency--phase portrait is a two-dimensional array $S(\nu_i, \phi_j)$ of intensity as a function of frequency channel $\nu_i$ and pulse phase bin $\phi_j$, replicating the observational setup of the I-LOFAR telescope. This data product was chosen for its ability to discern key features of profile evolution, such as the dispersion sweep or changing ToAs, directly.

The simulator operates on a grid of $N_\nu = 40$ linearly spaced frequency channels spanning $112.01$--$190.14$ MHz, with a reference frequency of $\nu_\mathrm{ref} = 151.07$ MHz. We initially trained the network using a simulator with logarithmically spaced channels, which offer finer resolution at the low-frequency end of the band, where dispersive delay is steepest. Real archives, however, are reduced to $N_\nu = 40$ channels by linearly averaging (frequency-scrunching) the native-resolution data. We therefore adopted linear spacing to match this channelization, so that channel index corresponds to the same frequency in simulated and real portraits. The phase axis is divided into $N_\phi = 512$ bins over one full rotation period of $P_0 = 16.052$ ms (spin frequency $f = 62.296$ Hz for PSR~J2145$-$0750).

\subsection{Profile component model}\label{sec:Procompmodel}
Each profile component is described by five parameters, defined at a reference frequency $\nu_\mathrm{ref}$:

\begin{itemize}
    \item $A_\mathrm{ref}$: amplitude at $\nu_\mathrm{ref}$
    \item $\phi_0$: phase centre in the range $[0, 1)$
    \item $\kappa_\mathrm{ref}$: von Mises concentration parameter (larger values give a narrower component)
    \item $\gamma$: spectral index of the amplitude
    \item $\beta$: spectral index of the concentration (width evolution)
\end{itemize}

At each frequency channel $\nu_i$, the component amplitude and width are scaled as power laws in the frequency ratio $r_i = \nu_i / \nu_\mathrm{ref}$:

$$A(\nu_i) = A_\mathrm{ref} r_i^{\gamma},$$

$$\kappa(\nu_i) = \kappa_\mathrm{ref}r_i^{\beta}.$$

The intrinsic pulse shape at channel $i$ is a von Mises distribution:

$$S_\mathrm{comp}(\nu_i, \phi) = A(\nu_i) \exp\left[\kappa(\nu_i)\bigl(\cos(2\pi\Delta\phi) - 1\bigr)\right].$$

We choose this functional form because it is periodic on $[0, 1)$ by construction and uses a single parameter, $\kappa$, to control width. It also reduces to a Gaussian in the limit of large $\kappa$, appropriate for the narrow pulse profiles of millisecond pulsars. The clean portrait before dispersion, scattering, or noise are applied is the sum of contributions from all components:
$$S_\mathrm{clean}(\nu_i, \phi) = \sum_k S_\mathrm{comp}^{(k)}(\nu_i, \phi).$$

\begin{figure*}
    \centering
    \includegraphics[width=\linewidth]{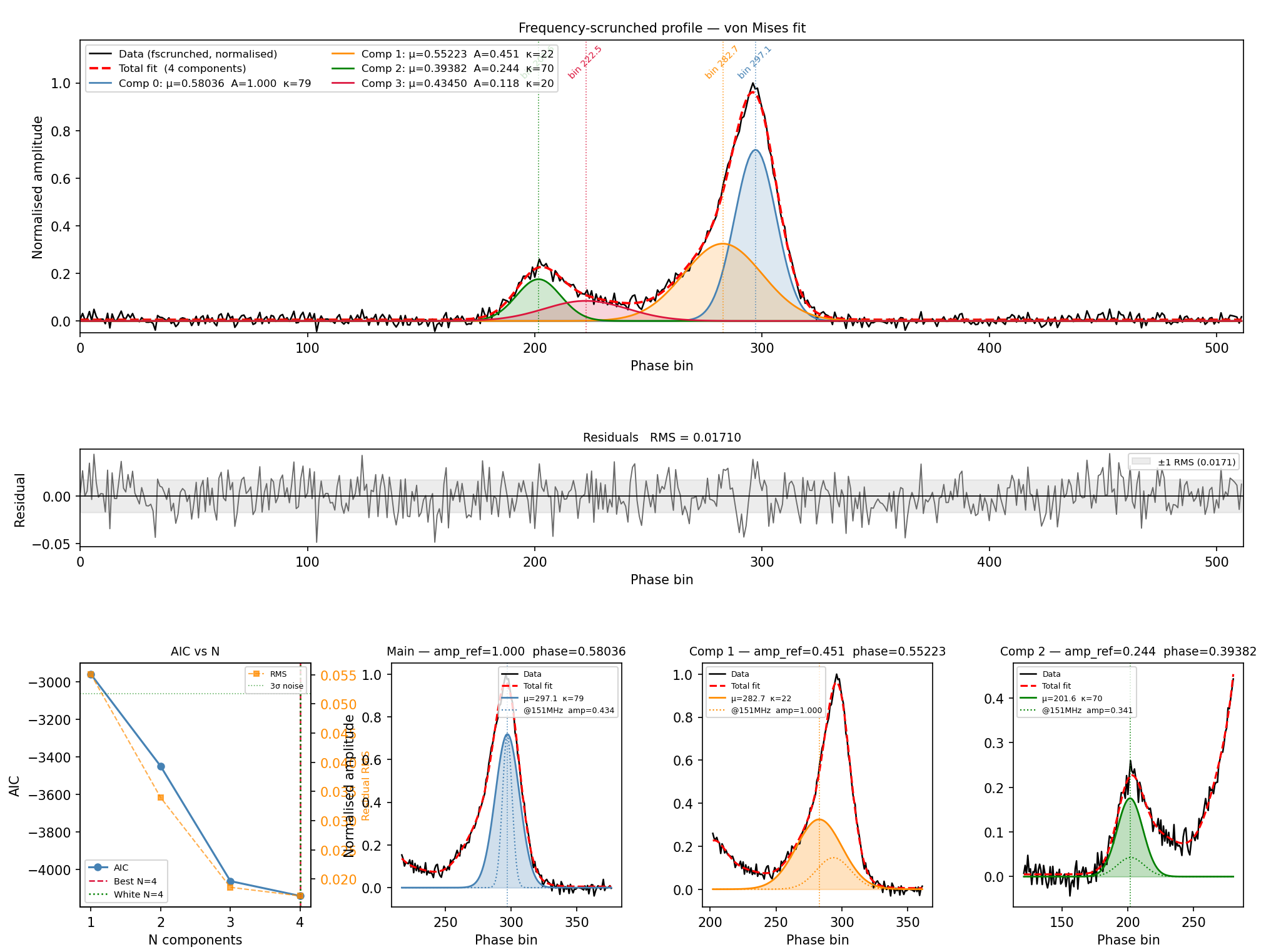}
    \caption{Significant profile components identified in a high-S/N I-LOFAR archive of PSR~J2145$-$0750. An AIC-based model comparison selects four components as the best-fitting model, reducing the fit residual to the noise floor.}
    \label{fig:vmfit}
\end{figure*}

\subsection{\texorpdfstring{PSR~J2145$-$0750 component geometry}{PSR J2145-0750 component geometry}}\label{sec:compgeom}
We decompose the profile into a small number of analytic components and attach the frequency evolution to those components individually. This has two consequences that matter for the inference. First, the profile at any frequency follows from the component parameters rather than from interpolation between measured channels, so the simulator extrapolates in a physically controlled way. Second, frequency evolution becomes a property of identifiable features rather than of the profile as a whole, which is what allows the amplitude spectral indices to be inferred at all.

We derive the component geometry from a single I-LOFAR observation of PSR~J2145$-$0750 taken on 2023 July 26, chosen for having the highest S/N of any epoch in our dataset. The number of components is determined from the data by fitting models with $N = 1, 2, 3, \dots$ von Mises components to the frequency-averaged profile, placing each successive component at the largest absolute residual peak furthest from any existing component, and evaluate each fit in two ways. The first is the Akaike information criterion, 
\begin{equation}
    \mathrm{AIC} = n\ln\!\left(\frac{\mathrm{RSS}}{n}\right) + 2k,
\end{equation}
where $n = 512$ is the number of phase bins, $\mathrm{RSS} = \sum_j \bigl[P(\phi_j) - \hat{P}(\phi_j)\bigr]^2$ is the residual sum of squares between the observed profile $P$ and the fitted model $\hat{P}$, and $k = 3N + 1$ is the number of free parameters: an amplitude, position and concentration for each of the $N$ components ($3N$), plus a constant baseline (+1). We require $\Delta\mathrm{AIC} < -2$ before an additional component is accepted guarding against components that reduce the residual by less than their cost in model complexity. The second, is a whiteness test on the residuals: we measure the off-pulse noise level $\sigma_\mathrm{off}$ from the bins below 5\% of the profile peak, and stop when $\max|\mathrm{residual}| < 3\sigma_\mathrm{off}$. At that point the residuals are statistically indistinguishable from the noise, and any further component would result in over-fitting. The whiteness criterion takes precedence over the AIC result, since a model whose residuals have reached the noise floor is complete regardless of whether AIC would tolerate more freedom.
\begin{table*}
\centering
\caption{Intrinsic parameters of the four von Mises components used to model the PSR~J2145$-$0750 pulse profile. Parameters marked U$[a,b]$ are drawn uniformly from the stated range at each simulation; fixed values are constant across all simulations. The global phase offset $\Delta\phi$ (the \texttt{toa} parameter) is added to every component phase centre and is inferred jointly with the other five SBI parameters. The amplitude and width of each component evolve as shown above with their respective spectral indices.}
\label{tab:components}
\renewcommand{\arraystretch}{1.35}
\begin{tabular}{lcccc}
\hline\hline
Parameter & Main & Pre-main bridge & Leading & Inter bridge \\
\hline
Phase centre $\phi_0$ (cycles)
    & $0.58036 + \Delta\phi$
    & $0.55223 + \Delta\phi$
    & $0.39382 + \Delta\phi$
    & $0.43450 + \Delta\phi$ \\
Reference amplitude $A_\mathrm{ref}$
    & $1.0$ (fixed)
    & $\mathcal{U}[0.25,\;0.65]$
    & $\mathcal{U}[0.10,\;0.50]$
    & $\mathcal{U}[0.05,\;0.25]$ \\
Reference concentration $\kappa_\mathrm{ref}$
    & $\mathcal{U}[40,\;100]$
    & $\mathcal{U}[10,\;40]$
    & $\mathcal{U}[40,\;100]$
    & $\mathcal{U}[8,\;30]$ \\
Amplitude spectral index $\gamma_A$
    & $\gamma_\mathrm{main}$
    & $\gamma_\mathrm{main}$
    & $\gamma_\mathrm{lead}$
    & $\gamma_\mathrm{lead}$ \\
Width spectral index $\gamma_\kappa$
    & $0.5$ (fixed)
    & $0.5$ (fixed)
    & $2.5$ (fixed)
    & $2.5$ (fixed) \\
\hline
\end{tabular}
\end{table*}
For PSR~J2145$-$0750 this procedure selects $N = 4$. The four components form two pairs: a trailing complex comprising the dominant $main$ component and a broader, weaker $pre$-$main$ bridge, and a leading complex comprising the $leading$ component and a broader $inter$ bridge. Their parameters are listed in Table~\ref{tab:components} and the best fit is shown in Fig.~\ref{fig:vmfit}. We assign a single amplitude spectral index to each pair, $\gamma_\mathrm{main}$ to the trailing complex and $\gamma_\mathrm{lead}$ to leading complex. This reflects the assumption that each pair arises from a common emission region. The component phase centres are held fixed at their fitted values in all simulations, with only the global offset $\Delta\phi$ varying, and this is a deliberate identifiability choice rather than a computational convenience.

\begin{figure*}[!h]
    \centering
    \includegraphics[width=\linewidth]{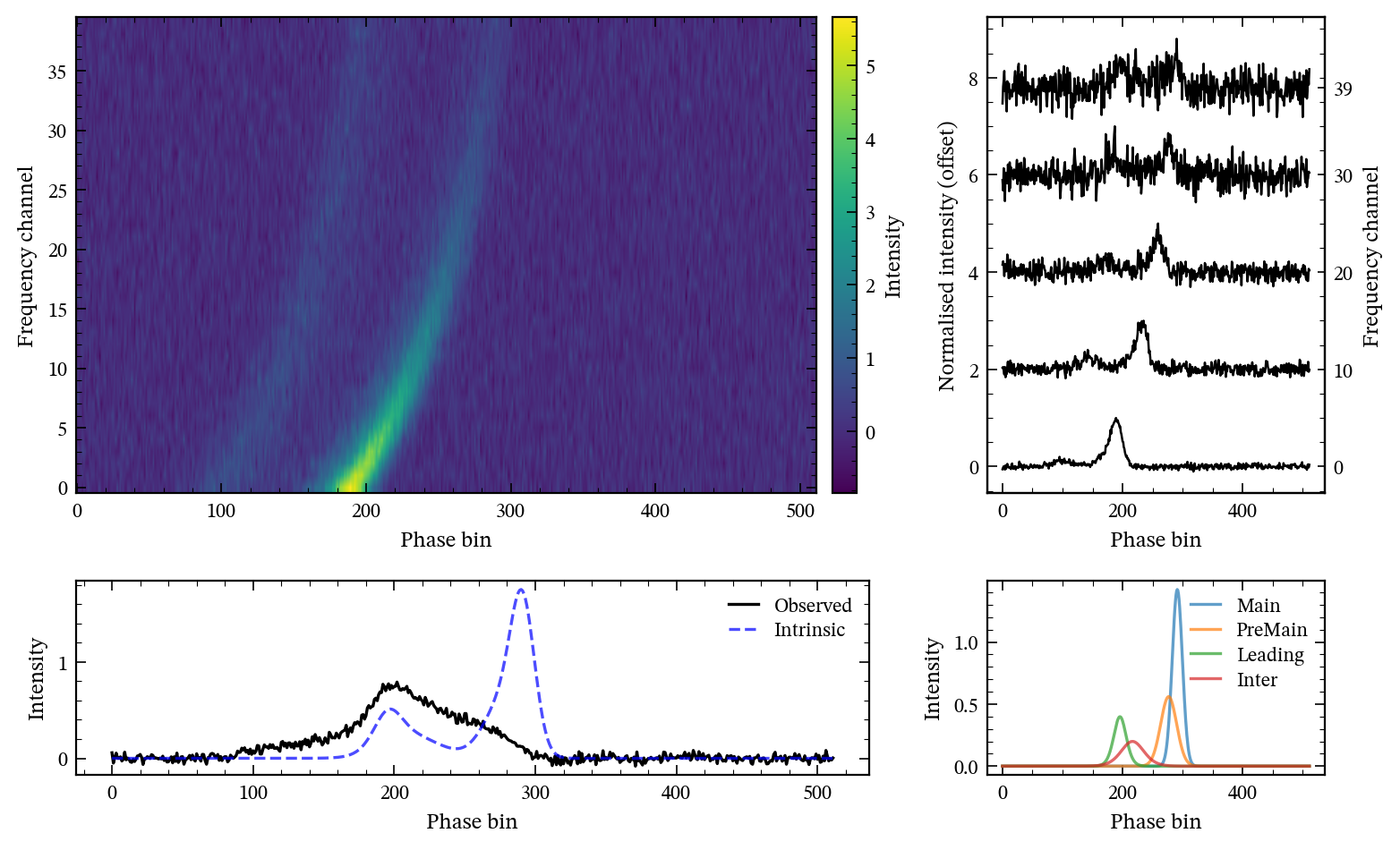}
    \caption{Example of a simulated portrait with values drawn from the priors detailed in Table~\ref{tab:parameters}. The top left panel shows the 40$\times$512 portrait. The left vertical panel shows the individual channel pulse profiles with the dispersive sweep following from that of the portrait. The bottom left panel shows both the observed and intrinsic pulse profiles whereas the bottom right panel shows the individual components. }
    \label{fig:portrait}
\end{figure*}

\subsection{Dispersion}\label{sec:dispersion}
The dispersive delay of a pulse at frequency $\nu$ relative to the highest frequency in the band, $\nu_\mathrm{max}$, is:

$$\Delta t(\nu) = \mathcal{D} \cdot \mathrm{DM} \left(\frac{1}{\nu^2} - \frac{1}{\nu_\mathrm{max}^2}\right),$$ 
where $\mathcal{D} = 4.1488 \times 10^6$ MHz$^2$ ms (pc~cm$^{-3}$)$^{-1}$ is the dispersion constant, and DM is the residual dispersion measure in pc~cm$^{-3}$ \citep{kulkarni2020}, drawn from $\mathcal{U}[-0.025, 0.025]$ pc cm$^{-3}$. This range reflects that PSR~J2145$-$0750's DM is already known to high precision from previous timing. In practice, real archives are first dedispersed at their reference DM before being passed to the SBI pipeline, which then infers only the small residual left over from that starting point. The delay is zero at the highest frequency of the band and grows in magnitude as frequency decreases. This delay in terms of phase bins will be:  

$$\Delta n(\nu_i) = \frac{\Delta t(\nu_i)}{P_0} \cdot N_\phi.$$

\noindent The delay is applied as a fractional circular shift in the Fourier domain. For each channel $i$, the shifted profile is:
$$\tilde{S}(\nu_i, \phi) = \mathcal{F}^{-1}\left[\mathcal{F}\{S_\mathrm{clean}(\nu_i, \phi)\} e^{-2\pi i k \Delta n(\nu_i) / N_\phi}\right],$$
where $k$ is the discrete Fourier mode index, $\mathcal{F}$ depicts the Fourier transform and is taken along the phase axis $\phi$. This follows directly from the Fourier shift theorem: a phase gradient in Fourier space is a translation in the phase domain. Performing the shift this way preserves the periodic boundary of the pulse phase exactly, avoids interpolation artefacts, and naturally accommodates sub-bin fractional delays.

\subsection{Noise}\label{sec:noise}
Noise is added to the simulations as independent Gaussian realisations per pixel, with a per-pixel standard deviation calibrated to the target S/N. The S/N is defined with respect to the peak of the frequency-averaged profile:

$$\sigma_\mathrm{pix} = \frac{A_\mathrm{avg,peak}}{\mathrm{S/N}} \sqrt{N_\nu},$$
where $A_\mathrm{avg,peak}$ is the peak amplitude of the noise-free frequency-averaged profile and the $\sqrt{N_\nu}$ factor accounts for the noise averaging down over channels. The S/N is drawn log-uniformly from $\mathcal{U}[\ln 7, \ln 120]$ and is parametrised as $\log(\mathrm{S/N})$ in the inference. The resulting portrait is:

$$S(\nu_i, \phi_j) = \tilde{S}(\nu_i, \phi_j) + \mathcal{N}(0, \sigma_\mathrm{pix}).$$
A fixed, low-level channel-to-channel bandpass irregularity is also applied: a multiplicative log-normal gain per channel with $\sigma_\mathrm{bp}=0.1$, redrawn independently for each simulation. This is a marginalised systematic and not a free parameter of inference.

\begin{table*}
\centering
\caption{Complete parameter inventory of the J2145$-$0750 forward simulator.
\textit{Free} parameters form the six-dimensional inference target $\boldsymbol{\theta}$. \textit{Marginalised} parameters are drawn independently for each simulation and implicitly integrated out during training, so the posterior is robust to their variation. \textit{Fixed} parameters are constants set from a fit to a real reference observation (component phase centres; see Table~\ref{tab:components}), prior testing, or the observational setup.}
\label{tab:parameters}
\renewcommand{\arraystretch}{1.3}
\begin{tabular}{llcc}
\hline\hline
Parameter & Symbol & Prior / Value & Status \\
\hline
\multicolumn{4}{l}{\textit{\textbf{Inference parameters}}} \\[2pt]
DM residual                              & $\Delta\mathrm{DM}$   & $\mathcal{U}[-0.025,\;0.025]$ pc\,cm$^{-3}$ & Free \\
Main amplitude spectral index            & $\gamma_\mathrm{main}$  & $\mathcal{U}[-6,\;1]$                         & Free \\
Leading amplitude spectral index         & $\gamma_\mathrm{lead}$  & $\mathcal{U}[-4,\;1]$                         & Free \\
Phase offset (ToA)                       & $\Delta\phi$            & $\mathcal{U}[-0.08,\;0.08]$ cycles            & Free \\
Log signal-to-noise ratio                & $\ln(\mathrm S/N)$     & $\mathcal{U}[\ln 7,\;\ln 120]$                & Free \\
Leading reference amplitude              & $A_{ref}$       & $\mathcal{U}[0.10,\;0.50]$                    & Free \\[4pt]
\multicolumn{4}{l}{\textit{\textbf{Nuisance parameters (marginalised)}}} \\[2pt]
Main reference concentration             & $\kappa_\mathrm{main}$    & $\mathcal{U}[40,\;100]$                       & Marginalised \\
Pre-main reference concentration         & $\kappa_\mathrm{premain}$ & $\mathcal{U}[10,\;40]$                        & Marginalised \\
Leading reference concentration          & $\kappa_\mathrm{lead}$    & $\mathcal{U}[40,\;100]$                       & Marginalised \\
Inter-bridge reference concentration     & $\kappa_\mathrm{inter}$   & $\mathcal{U}[8,\;30]$                         & Marginalised \\
Pre-main reference amplitude             & $A_\mathrm{premain}$      & $\mathcal{U}[0.25,\;0.65]$                    & Marginalised \\
Inter-bridge reference amplitude         & $A_\mathrm{inter}$        & $\mathcal{U}[0.05,\;0.25]$                    & Marginalised \\
Per-channel bandpass factor ($\times$40) & $b_i$                     & $\mathrm{LogNormal}(0,\;0.1)$                 & Marginalised \\[4pt]
\multicolumn{4}{l}{\textit{\textbf{Fixed parameters}}} \\[2pt]
Main reference amplitude                 & $A_\mathrm{main}$         & $1.0$                                         & Fixed \\
Component phase centres (4)              & $\phi_{0,k}$              & see Table~\ref{tab:components}               & Fixed \\
Width spectral index (main, pre-main)    & $\gamma_{\kappa,\mathrm{main}}$ & $0.5$                                  & Fixed \\
Width spectral index (leading, inter)    & $\gamma_{\kappa,\mathrm{lead}}$  & $2.5$                                  & Fixed \\
Phase-drift spectral index (all)         & $\gamma_{\phi}$             & $0.0$                                         & Fixed \\
Scattering timescale                     & $\tau_\mathrm{scatter}$     & $0.0$ ms                                      & Fixed \\
Bandpass irregularity $\sigma$           & $\sigma_\mathrm{bp}$        & $0.1$                                         & Fixed \\
Reference frequency                      & $\nu_\mathrm{ref}$          & $151.07$ MHz                                  & Fixed \\
\hline
\end{tabular}
\end{table*}

\subsection{Scattering}\label{sec:scat}
Scatter broadening is applied channel-by-channel as a convolution of the intrinsic profile with a one-sided exponential kernel. The scattering timescale at the reference frequency, $\tau_\mathrm{ref}$, is scaled to each channel with a power-law frequency dependence, $\tau(\nu_i) = \tau_\mathrm{ref} (\nu_i / \nu_\mathrm{ref})^\alpha$, where $\alpha$ is the scattering spectral index (typically $-4$ for a thin screen, $-4.4$ for Kolmogorov turbulence). In the current model, $\tau_\mathrm{ref}$ is fixed at zero: the expected scattering timescale for PSR~J2145$-$0750 at 150 MHz is sub-bin ($\tau \lesssim 10\,\mu$s against a bin width of $31.4\,\mu$s), and all real-data maximum likelihood estimates yield $\tau \to 0$. We also note that we can also simulate the presence of scintillation in the portraits but in this case, we ignored it as it is negligible for this pulsar at these frequency bands. Fig~\ref{fig:portrait} shows an example of a simulated portrait with values drawn from the priors detailed in Table~\ref{tab:parameters}. This portrait is built after including all the effects described above.

\begin{figure*}
    \centering
    \includegraphics[width=\linewidth]{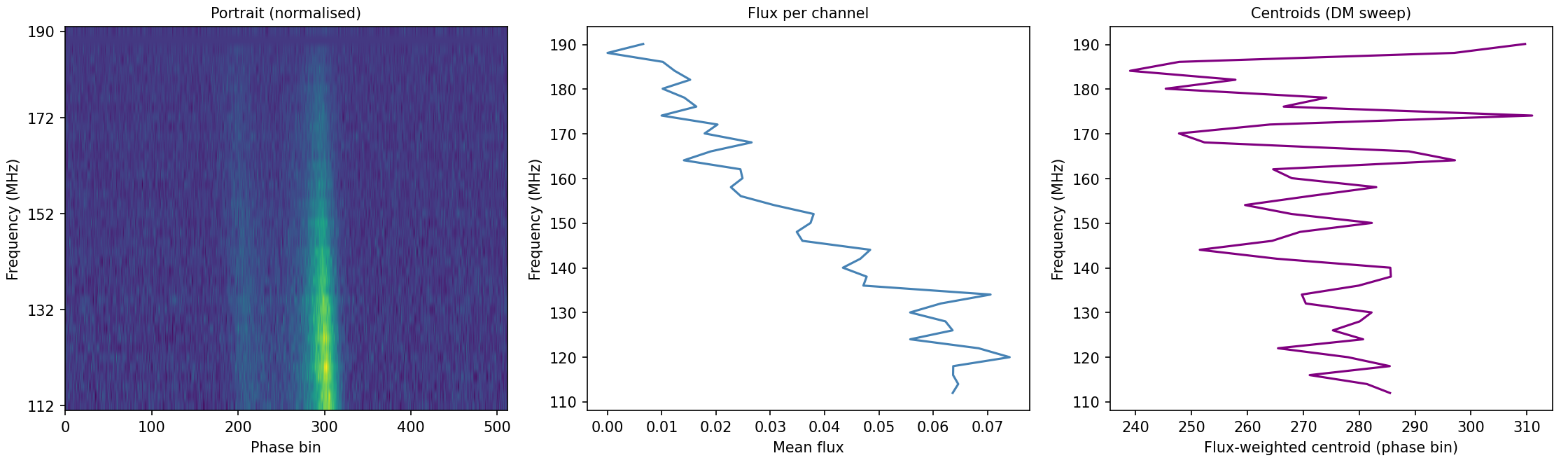}
    \caption{The data products extracted from a single real I-LOFAR observation of PSR~J2145$-$0750. \textit{First panel}: the normalised frequency--phase portrait, processed by the convolutional branch of the embedding network. \textit{Second}: the per-channel flux, which constrains the amplitude spectral indices. \textit{Third}: the per-channel centroids, whose gradient carries the residual dispersion measure and whose overall level carries the arrival time.}
    \label{fig:sim_dia}
\end{figure*}

\subsection{Data products to perform SBI}\label{sec:dataprod}

The observation vector fed to the posterior estimator consists of the normalised portrait together with two per-channel summary statistics. Both of them are deterministic functions of the portrait itself, so they add no information that is not already present in $S(\nu_i, \phi_j)$. Their purpose is representational: a CNN can in principle learn any such statistic from the portrait, but with a fixed architecture and a finite training budget it must first discover it. Supplying the statistics directly makes the structure that constrains each parameter immediately accessible, and we found this to be necessary in practice rather than merely convenient.

We arrived at this feature set incrementally, adding a branch whenever a newly introduced parameter proved poorly constrained by the existing inputs. With the portrait alone, $\Delta$DM and $\log(\mathrm{S/N})$ were recovered well, but introducing the amplitude spectral index $\gamma_\mathrm{main}$ degraded the inference substantially. Adding the per-channel flux resolved this. Introducing the arrival time then produced a degeneracy with $\Delta$DM, since both displace the pulse in phase; adding the per-channel centroids resolved that in turn. The remaining two parameters required no further inputs, and the three-branch structure (including CNN) was sufficient for the full six-parameter model. The two summary statistics branches are as follows:
\begin{itemize}
    \item \textbf{Per-channel flux} ($N_\nu = 40$): the mean intensity across all phase bins in each channel, $\bar{S}_i = \langle S(\nu_i, \phi)\rangle$. This is the band spectrum of the pulse, and it constrains the amplitude spectral indices directly: since component amplitudes scale as $(\nu/\nu_\mathrm{ref})^{\gamma_A}$, the run of $\bar{S}_i$ with frequency is a $40$-point sampling of that power law. The same information is present in the portrait, but distributed across all $N_\phi$ phase bins of each channel; collapsing the phase axis concentrates it into the form in which it bears on $\gamma_\mathrm{main}$ and $\gamma_\mathrm{lead}$.

    \item \textbf{Per-channel centroids} ($N_\nu = 40$): the flux-weighted mean phase bin in each channel,
    \begin{equation}
        c_i = \frac{\sum_j n_j\, S(\nu_i, n_j)}{\sum_j S(\nu_i, n_j)},
    \end{equation} 
    where $n_j$ indexes phase bins. The centroids separate the two effects that displace the pulse in phase. A residual dispersion measure shifts channel $i$ by an amount $\propto \nu_i^{-2}$, imposing a characteristic curvature on $c_i$. Subsequently, a rigid ToA offset shifts every channel equally and therefore changes only the overall level of the curve. The gradient thus carries $\Delta$DM while the level carries the ToA. 
\end{itemize}
The full observation vector is the concatenation
\begin{equation}
    \mathbf{x} = \Bigl[
        \mathrm{vec}(S),\;
        \{\bar{S}_i\}_{i=1}^{N_\nu},\;
        \{c_i\}_{i=1}^{N_\nu}
    \Bigr] \in \mathbb{R}^{20{,}560},
\end{equation}
where $\mathrm{vec}(S) \in \mathbb{R}^{N_\nu N_\phi} = \mathbb{R}^{20{,}480}$ is the flattening of the portrait, adding the other two branches gives the final dimensionality of $20{,}480 + 40 + 40 = 20{,}560$.

A fourth statistic, the frequency-averaged mean profile $\bar{P}(\phi_j) = N_\nu^{-1}\sum_i S(\nu_i, \phi_j)$, was introduced when we tested the inclusion of interstellar scattering as a free parameter. Averaging over channels improves the profile S/N by $\sqrt{N_\nu} \approx 6.3$, which makes the asymmetric exponential tail characteristic of scattering more readily distinguishable from the symmetric dispersive sweep than it is in any single channel. Since it was excluded from the final model (Sect.~\ref{sec:scat}), this branch was removed. We retain the result as a general principle: because every branch is a function of the portrait, the feature set can be extended to expose whatever structure a newly introduced parameter requires. We show the three data products used here, for a single real observation in Fig.~\ref{fig:sim_dia}.

\subsection{Neural Posterior Estimation}\label{sec:npe}
Neural posterior estimation (NPE) is an SBI technique that trains a conditional density estimator $q(\theta|x)$ to approximate the Bayesian posterior $p(\theta|x)$, using pairs $\{\theta, x\}$ generated by the forward simulator. In our case, the simulator also depends on a set of nuisance parameters that are not part of the inference target:
$$\mathbf{x} = f(\boldsymbol{\theta}, \boldsymbol{\nu}),$$
where $\boldsymbol{\nu}$ comprises six scalar nuisance parameters (four component concentrations, two inter-component amplitudes) drawn from their own priors at each simulation, plus a further 40 per-channel bandpass factors (Sect.~\ref{sec:noise}). All of these are implicitly marginalised over during training, so the network learns a posterior already integrated over profile-shape and bandpass variability.

We use a neural spline flow (NSF) as the density estimator, using piecewise-polynomial rational-quadratic splines as flexible, invertible transformations. The flow is trained by maximum likelihood on simulated $\{\theta, x\}$ pairs, using the \texttt{SNPE} implementation \citep{Papa2016} from the \texttt{sbi} package~\citep{tejero2020} in its single-round, fully amortized configuration. The proposal distribution is always the prior, so the trained posterior is valid for any observation consistent with the prior and the simulator, without further simulation or retraining at inference time. The embedding network is divided into three branches, each targeting a different aspect of the portrait, whose outputs are concatenated and passed through a final combination layer (Table~\ref{tab:encoder}):
\begin{itemize}
\item \textbf{Portrait branch:} a 2D convolutional neural network (CNN) capturing the joint morphology of the portrait, primarily responsible for the DM sweep, pulse widths, and amplitude spectral indices.

\item \textbf{Flux branch:} a multi-layer perceptron (MLP) on the per-channel flux, encoding the broadband spectral shape that constrains $\gamma_\mathrm{main}$ and $\gamma_\mathrm{lead}$ independently of profile structure.

\item \textbf{Centroid branch:} an MLP on the per-channel centroids (concatenated with the portrait branch's output), separating the phase shift due to DM from that due to ToA and helping resolve the degeneracy between them.
\end{itemize}

\subsection{Embedding network}\label{sec:embedding}
The final observation vector has 20,560 dimensions (Sect.~\ref{sec:dataprod}) and is highly redundant. The NSF requires a compressed summary retaining only the information relevant to each parameter. Table~\ref{tab:encoder} gives the full architecture: an NSF with hidden width 64 and 5 spline components, conditioned on a 128-dimensional embedding produced by the three branches above.

\begin{table*}
\centering
\caption{Multi-branch Encoder architecture. All activations are ReLU.
         The centroid branch input concatenates the 40 centroid values
         with the 128-dimensional portrait CNN output.
         The final summary vector $\mathbf{z}$ is 128-dimensional.}
\label{tab:encoder}
\setlength{\tabcolsep}{5pt}
\begin{tabular}{lllll}
\toprule
\textbf{Branch} & \textbf{Layer} & \textbf{Config} & \textbf{Input shape} & \textbf{Output shape} \\
\midrule
\multirow{9}{*}{Portrait CNN}
 & Conv2d            & $3{\times}3$, 8 filters  & $1\times40\times512$  & $8\times40\times512$  \\
 & MaxPool2d         & $2{\times}2$             & $8\times40\times512$  & $8\times20\times256$  \\
 & Conv2d            & $3{\times}3$, 16 filters & $8\times20\times256$  & $16\times20\times256$ \\
 & MaxPool2d         & $2{\times}2$             & $16\times20\times256$ & $16\times10\times128$ \\
 & Conv2d            & $3{\times}3$, 32 filters & $16\times10\times128$ & $32\times10\times128$ \\
 & MaxPool2d         & $2{\times}2$             & $32\times10\times128$ & $32\times5\times64$   \\
 & Conv2d            & $3{\times}3$, 32 filters & $32\times5\times64$   & $32\times5\times64$   \\
 & AdaptiveAvgPool2d & $(5,4)$                  & $32\times5\times64$   & $32\times5\times4$    \\
 & Linear            & $640\to128$              & $640$                 & $128$                 \\
\midrule
\multirow{2}{*}{Flux MLP}
 & Linear & $40\to64$  & $40$ & $64$ \\
 & Linear & $64\to32$  & $64$ & $32$ \\
\midrule
\multirow{2}{*}{Centroid MLP}
 & Linear & $168\to64$ & $168$ & $64$ \\
 & Linear & $64\to32$  & $64$  & $32$ \\
\midrule
\multirow{2}{*}{Combined FC}
 & Concatenate & $128+32+32$ & $128,\,32,\,32$ & $192$          \\
 & Linear      & $192\to128$ & $192$           & $\mathbf{128}$ \\
\bottomrule
\end{tabular}
\end{table*}

\subsection{Dataset generation \& training}\label{sec:dataset}
\begin{figure}
    \centering
    \includegraphics[width=\linewidth]{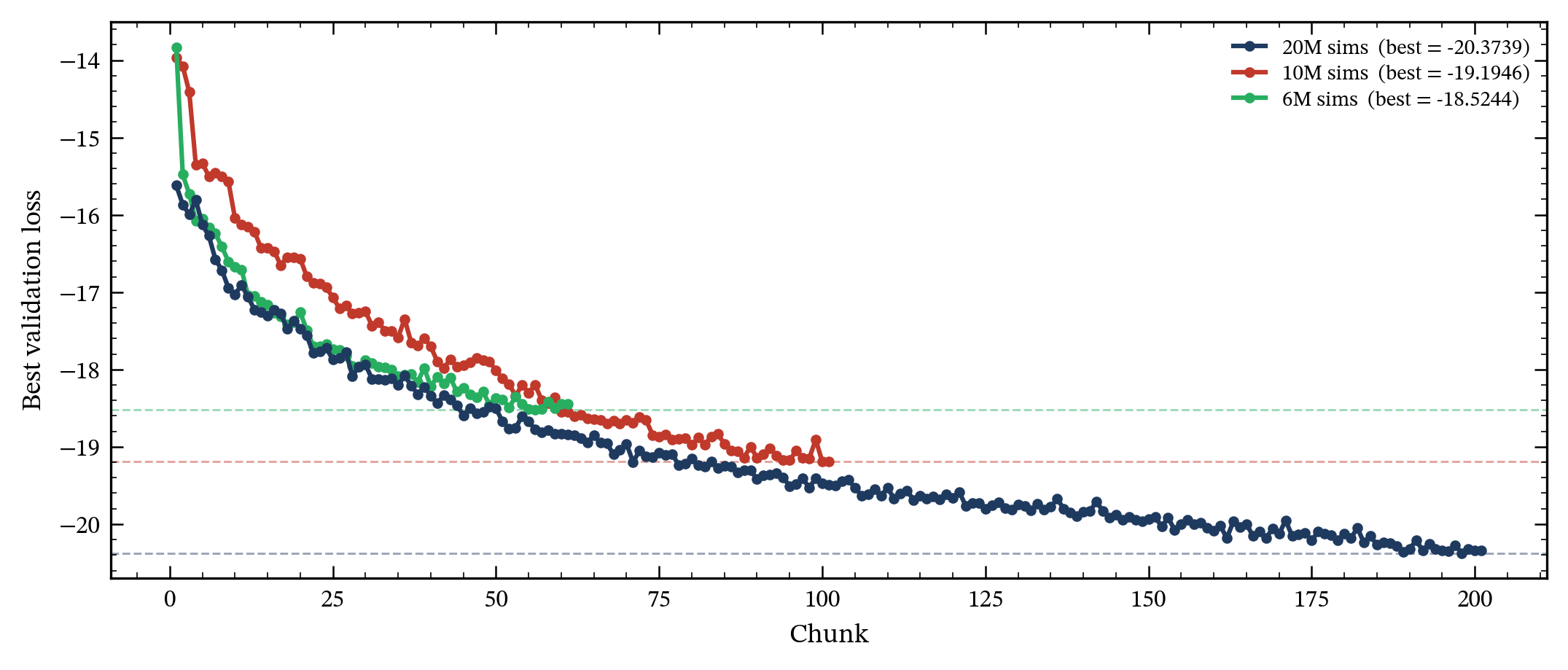}
    \caption{Validation loss for different simulations. The horizontal axis has chunk numbers with each chunk containing 10$^{5}$ simulations. The vertical axis is the validation loss. Increase in the number of simulations minimises the validation loss.}
    \label{fig:val_loss}
\end{figure}

Simulations are generated in parallel across 32 worker processes (16 physical CPU cores, hyper-threaded to 32 logical threads), streamed to disk in chunks of $10^5$ simulations each, giving $N_\mathrm{chunk} = N_\mathrm{sim}/10^5$ chunks. Generation proceeds approximately at the rate of $9.3\times10^5$ simulations per hour, so the cost scales linearly with $N_\mathrm{sim}$. Training then proceeds sequentially over the chunks. Each chunk is loaded into memory in turn and the neural spline flow is trained on it by maximum likelihood until early stopping (30 epochs without improvement in validation loss) or a per-chunk epoch cap, whichever comes first. Network weights are warm-started from the previous chunk's trained state, so the model accumulates information across the full simulation budget while never holding more than $10^5$ simulations in memory at once. All runs used a single NVIDIA L40S GPU (47.7 GB VRAM). With every run, the number of epochs required per chunk falls sharply after the first. Fig.~\ref{fig:val_loss} compares training outcomes at $6\times10^6$, $1\times10^7$, and $2\times10^7$ total simulations, with best validation losses of $-18.52$, $-19.19$, and $-20.37$ respectively. Validation loss continues to improve with training-set size across this entire range, with no indication of saturation at the largest budget tested. Both the $6\times10^6\to1\times10^7$ gain ($0.67$) and the $1\times10^7\to2\times10^7$ gain ($1.18$) exceed the run-to-run scatter by factors of $2$--$4$, indicating the improvement with additional training data is real rather than noise. Transfer efficiency which is defined as the ratio of epochs required by the first chunk to the last increase with the number of chunks trained, from $10.0\times$ at $6\times10^6$ simulations to $14.5\times$ at $2\times10^7$, consistent with the network requiring progressively less adaptation per chunk as training progresses.

\section{Validation \& Calibration}\label{sec:valcal}
In this section, we validate and test the statistical calibration of the trained model in two stages. First, compare the recovered and injected values of all six inferred parameters on independent simulations. Second, we test the statistical calibration of the posteriors by perfoming Probability-Probability (P-P) tests on them.

\begin{figure}
\centering
\includegraphics[width=\linewidth]{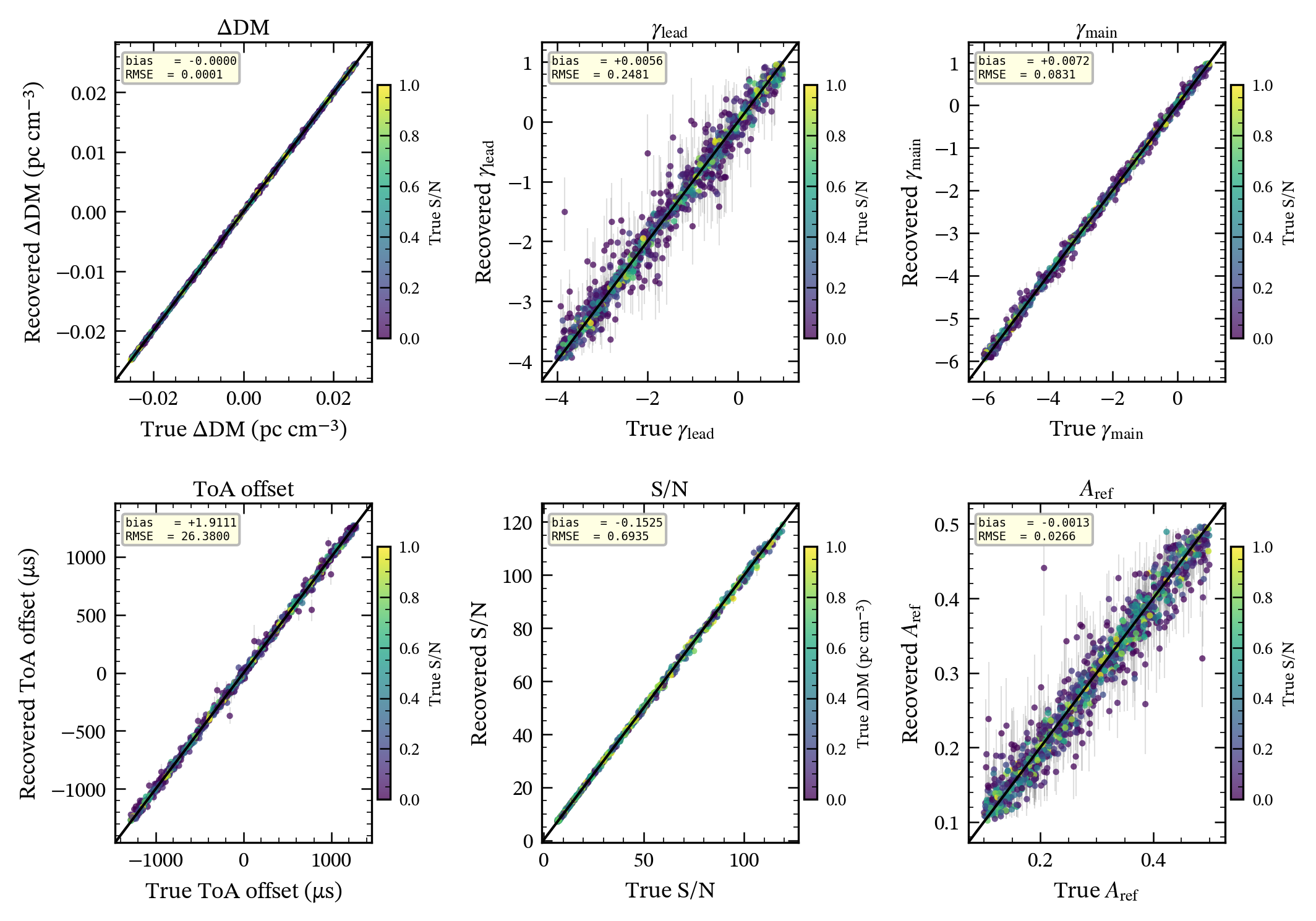}
\caption{Recovery of the six inferred parameters on 1000 independent simulated observations drawn from the prior. Each point compares the recovered posterior median with the injected true value, with error bars showing the $1\sigma$ posterior width and colour indicating the true S/N (except the S/N panel where the colour indicates true $\Delta DM$). The solid line marks perfect recovery.}
\label{fig:truevsrec}
\end{figure}

\subsection{Parameter recovery on simulated data}\label{sec:simul_valid}
To validate our method on simulated data, we assess the accuracy and precision of individual parameter recovery on 1000 independent simulations. For each, we infer the posterior median ($\hat\theta$) to compare with the true value ($\theta_\mathrm{true}$) and also compute the normalised residual, $(\hat{\theta}-\theta_\mathrm{true})/\sigma_{\hat\theta}$, where $\sigma_{\hat\theta}$ is its standard deviation. This measures accuracy and uncertainty scale simultaneously.

\begin{figure*}
\centering
\includegraphics[width=\linewidth]{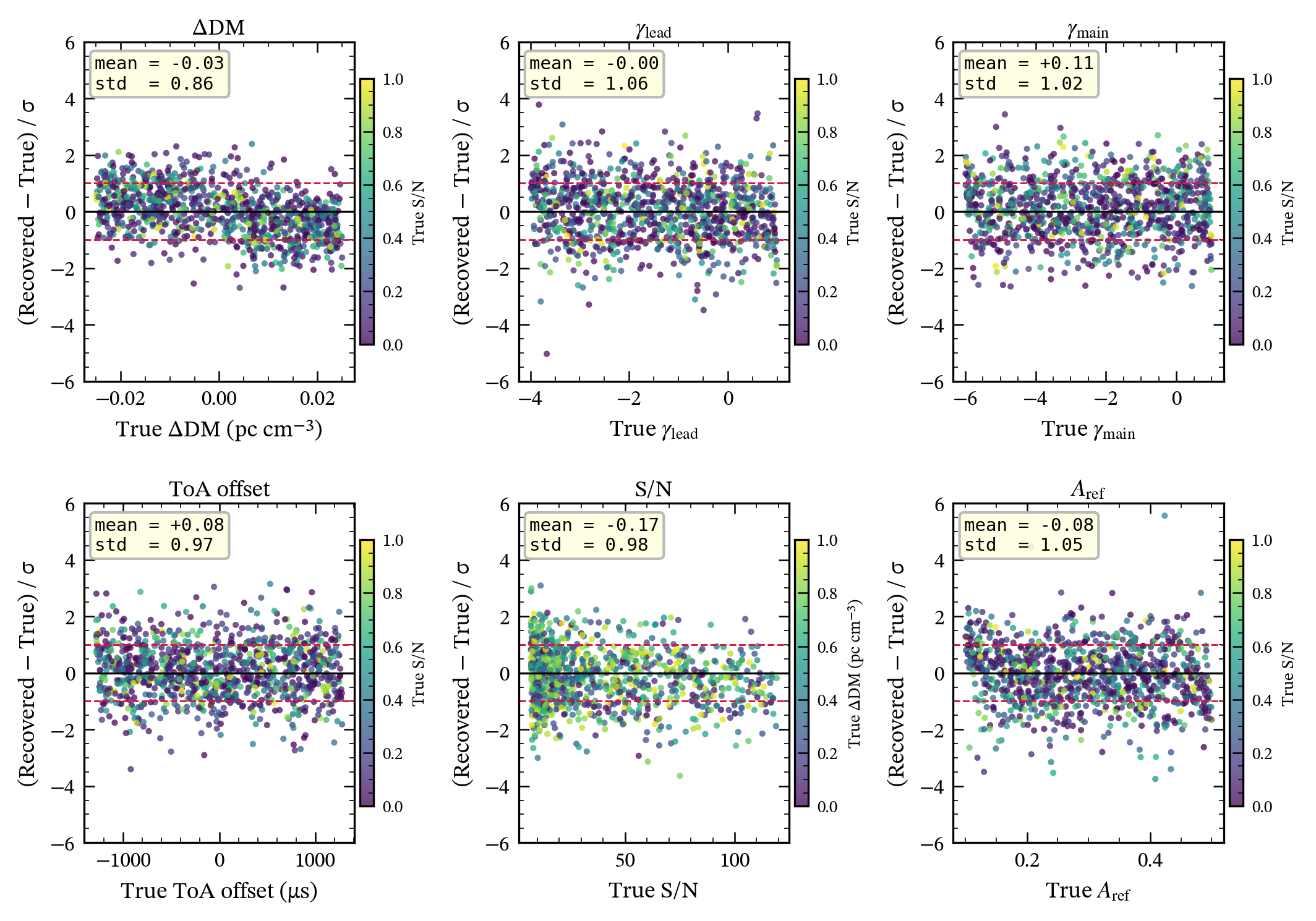}
\caption{Normalised residuals of the six inferred parameters on 1000 independent simulated observations drawn from the prior. Each point compares the amount of deviation in the recovered value with respect to the true value, and colour indicates the true S/N except for the bottom centre panel where the colour indicates true $\Delta DM$. There is no persistent bias seen in any parameter.}
\label{fig:normresid}
\end{figure*}

Fig.~\ref{fig:truevsrec} shows the recovered posterior medians against the injected true values for all six parameters, over 1000 independent simulations drawn from the prior. The recovery is unbiased across the full prior range: the DM and ToA are recovered with negligible bias and an RMSE of $1\times10^{-4}$~pc~cm$^{-3}$ and $26~\mu$s respectively. The profile-evolution indices are also well recovered, with the main-component index $\gamma_{\rm main}$ more tightly constrained than the weaker leading-component index $\gamma_{\rm leading}$. This is expected, as the leading component contributes less to the profile and is therefore harder to constrain. The precision degrades gradually towards low S/N, most visibly for A$_{ref}$, but with no catastrophic failures anywhere in the test set. Fig.~\ref{fig:normresid} shows the normalised residuals. Five of the six parameters have normalised standard deviations within 6\% of unity ($0.97$--$1.06$), indicating that the posterior widths are correctly sized and no parameter is significantly overconfident. The exception is $\Delta$DM at $0.86$, whose posteriors are approximately 16\% wider than required. Mean offsets are small throughout, the largest being $-0.17\sigma$ for $\log(\mathrm{S/N})$ and $+0.11\sigma$ for $\gamma_\mathrm{main}$. In absolute terms, the biases correspond to $0.0072$ in $\gamma_\mathrm{main}$ and $0.15$ in S/N, or 9\% and 22\% of the respective RMS error (RMSE).

\begin{figure}
\centering
\includegraphics[width=\linewidth]{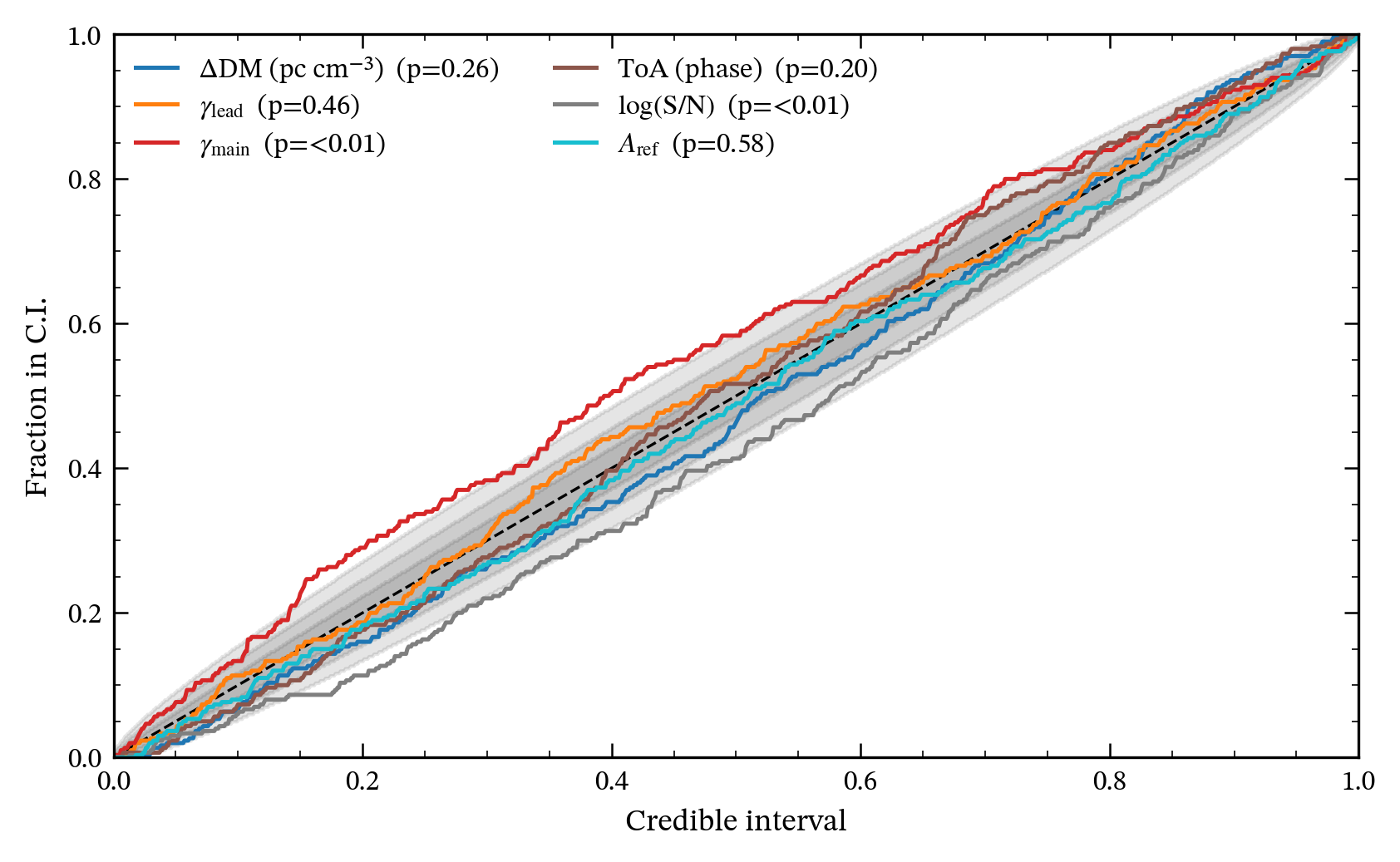}
\caption{P--P test of the SBI posteriors from 300 unseen simulated observations drawn from the prior. For each parameter, the curve shows the fraction of test cases for which the true value falls within a given marginal credible interval, plotted against that interval's nominal level. A perfectly calibrated posterior follows the diagonal (dashed line), and the shaded bands show the $1$, $2$ and $3\sigma$ ranges expected from the finite sample size. Legend values give the Kolmogorov--Smirnov $p$-value against uniformity for each parameter.}
\label{fig:pptest}
\end{figure}

Across the same 1000 simulations, the recovery residuals of $\Delta$DM and ToA are anti-correlated, with a Pearson coefficient ($r$) of $-0.36$, as expected since a change in DM can be partially absorbed by a shift in arrival time. This is the strongest pairwise correlation among the six parameters; the next largest, between $\gamma_\mathrm{lead}$ and $A_\mathrm{ref}$, is $+0.24$, and all remaining pairs satisfy $|r|<0.16$. The profile-evolution indices are therefore recovered largely independently of the timing parameters, showing that portrait-level inference separates the two efficiently.

\subsection{Statistical calibration of trained posteriors}
After demonstrating the recovery for the simulated data sets, we next verify that the posteriors are statistically well calibrated. The primary failure mode in SBI is the overconfidence of the posterior, which must be excluded explicitly. We perform P–P tests \citep{Cook2006} on an independent set of 300 simulated observations drawn from the prior. For each test case, we note the credible level at which the true value of each parameter falls within its marginal posterior, and compare the empirical distribution of these levels against the uniform distribution expected under perfect calibration. 

\begin{figure*}[!hb]
\centering
\includegraphics[width=\linewidth]{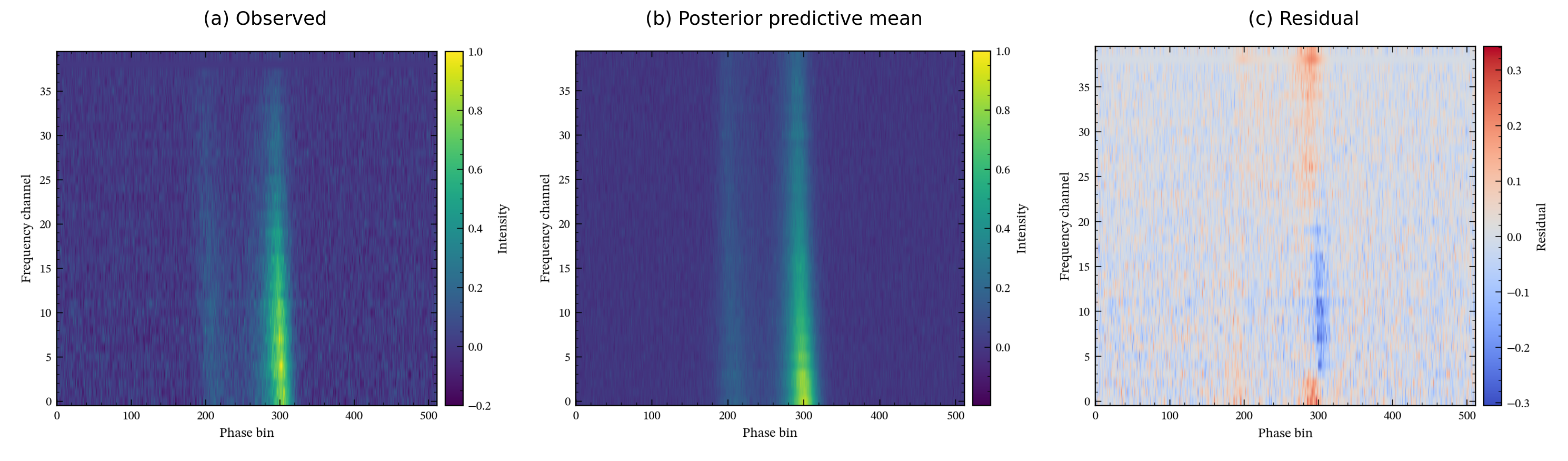}
\caption{Posterior-predictive check on a real I-LOFAR observation of PSR~J2145-0750. Left: observed frequency--phase portrait. Middle: 100 portraits generated from draws of the observation's posterior, overlaid; the reduced noise level relative to the observation reflects the averaging of independent noise realisations. Right: residual (observed minus posterior-predictive mean). Low-amplitude coherent structure confined to the on-pulse region ($\leq25\%$ of the pulse amplitude) reflects profile-shape detail beyond the two-component parametrisation, marginalised in the present model; the band in the uppermost channels arises from band-edge instrumental effects.}
\label{fig:ppc_rd}
\end{figure*}

Figure~\ref{fig:pptest} shows the resulting P–P curves for all six inferred parameters, together with an uncertainty estimate for the noise around the diagonal expected under Poisson counting error. At least four parameters, namely, $\gamma_\mathrm{leading}$, $A_\mathrm{ref}$, $\Delta$DM and ToA show no signs of bias (p-values greater than 0.01). We note that the sensitivity of the P--P test increases with the number of test cases, so the deviations that are significant at $N=300$ would not register at smaller $N$. We regard this as consistent with sampling scatter around good calibration rather than a substantive problem.
However, $\gamma_\mathrm{main}$ and $\log(\mathrm{S/N})$ show larger, statistically significant deviations ($p<0.01$ for both) in the current training run. These same parameters are shown to have larger mean offsets in Sect.~\ref{sec:simul_valid}. We attribute this to a small consistent shift in the posterior median rather than a wrongly estimated width. Repeating this test across six independent training runs spanning 6 million to 20 million simulations, we find that both of these parameters pass (p-values greater than 0.01) comfortably in half of them and fail in the other half. We note that no run shows a consistent directional bias across repeats. This pattern, along with similar rotation among $\Delta$DM, $\gamma_\mathrm{leading}$, ToA, and A$_{ref}$, each of which fails in some runs and passes comfortably in others, is consistent with training-realisation stochasticity rather than a systematic feature of any one parameter \citep{Alvey2026}.
We therefore conclude that the SBI approach is reasonably calibrated, but note that before wide-scale adoption this stochasticity would need to be addressed.

\section{Results}\label{sec:results}
\subsection{Application to I-LOFAR observations}
\label{sec:methods-data}

The I-LOFAR observations of PSR~J2145$-$0750 used in this work were reduced by Susarla et al (2025) \citep{susarla2025}, which describe RFI excision, polarisation calibration, and folding into archives of 512 phase bins and 400 frequency channels across the LOFAR HBA band. We take these calibrated archives as our starting point and do not repeat this reduction \citep{susarla_2025_dr}. We reduce the number of frequency channels from the native resolution to $N_\nu = 40$, both to keep network training computationally tractable and to match the frequency resolution produced by scrunching the real archives to the same channel count. The resulting portrait is $40\times512$ pixels, capturing the relevant propagation and profile-evolution effects for PSR~J2145$-$0750 at LOFAR frequencies.

\subsubsection{From archive to observation vector}
Each archive is validated against the training configuration before use. The phase-bin count must equal $N_\phi=512$, the channel count must be an exact integer multiple of $N_\nu=40$ (giving the frequency-scrunch factor, 10 for these data), and the band edges must fall within 2~MHz of the training range ($112.01$--$190.14$~MHz). Archives are then time-averaged to a single sub-integration, polarisation-scrunched to Stokes~$I$, baseline-subtracted, and dedispersed at the archive's own header DM. Because some archives may store channels in descending frequency order, we check the first and last channel's centre frequency and reverse the array where needed so that row 0 always corresponds to the lowest frequency, matching the simulator's convention.

\begin{figure*}[!h]
\centering
\includegraphics[width=\linewidth]{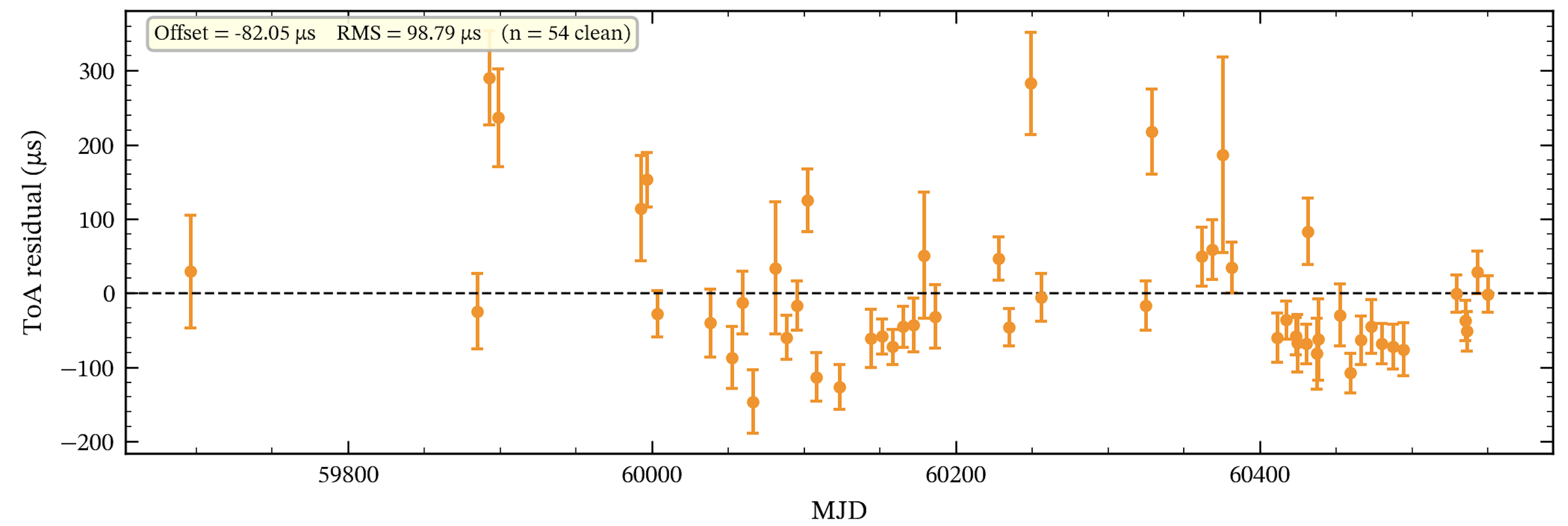}
\caption{Inferred ToA residuals for the I-LOFAR observations of PSR~J2145$-$0750 across the entire dataset with a constant offset of $-82.05~\mu$s removed, giving an RMS of $98.79~\mu$s over 54 ``clean'' epochs. Error bars show the $1\sigma$ posterior width in each case.}
\label{fig:timeseries}
\end{figure*}

The resulting $40\times512$ portrait is max-normalised, and the two summary features used in training are computed, namely the per-channel mean flux and the flux-weighted centroid in each channel (Sect.~\ref{sec:dataprod}). Real archives can contain channels with near-zero flux (RFI-flagged or otherwise dead), for which the centroid formula is numerically unstable and returns values far outside $[0, N_\phi)$. If a particular frequency channel falls outside the phase range, we zero the corresponding portrait row and flux entry, and set the centroid to a fixed neutral value (the phase bin of the Main component, Table~\ref{tab:components}), so that flagged channels contribute no information to the inference rather than propagating an unphysical value.

Before examining the inferred parameters, we verify that the forward model is describing the real data. Fig~\ref{fig:ppc_rd} compares the highest-S/N observed portrait with the mean of 100 portraits drawn from its posterior. The forward model reproduces the frequency evolution effects, with coherent residual structure confined to the on-pulse region and attributable to marginalisation of the nuisance parameters. We used the training model to get the posteriors on all the observations. Inference on a single observation took less than one second and the entire dataset of 56 epochs was processed in under a minute on a single GPU. The resulting posteriors are summarised by their median and $1\sigma$ credible intervals.

Fig.~\ref{fig:timeseries} shows the ToA residuals we infer for PSR~J2145$-$0750. We subtract a constant offset from the entire dataset to align the mean of the residuals to zero. This does not have any bearing towards the structure of ToA residuals. Two epochs are excluded: their recovered S/N fell at the lower edge of the prior, $\mathcal{U}[\ln 7, \ln 120]$, indicating a true S/N below the trained range. Since the network was never exposed to such observations, the ToA cannot be reliably inferred for them. We note that these residuals model the frequency evolution only, as the network is epoch-agnostic and does not include a time-domain timing model. The structure that remains between epochs is consequently unmodelled and it requires time domain analysis that lie outside the present scope. We defer it to future work.

\begin{figure*}[!ht]
\centering
\includegraphics[width=\linewidth]{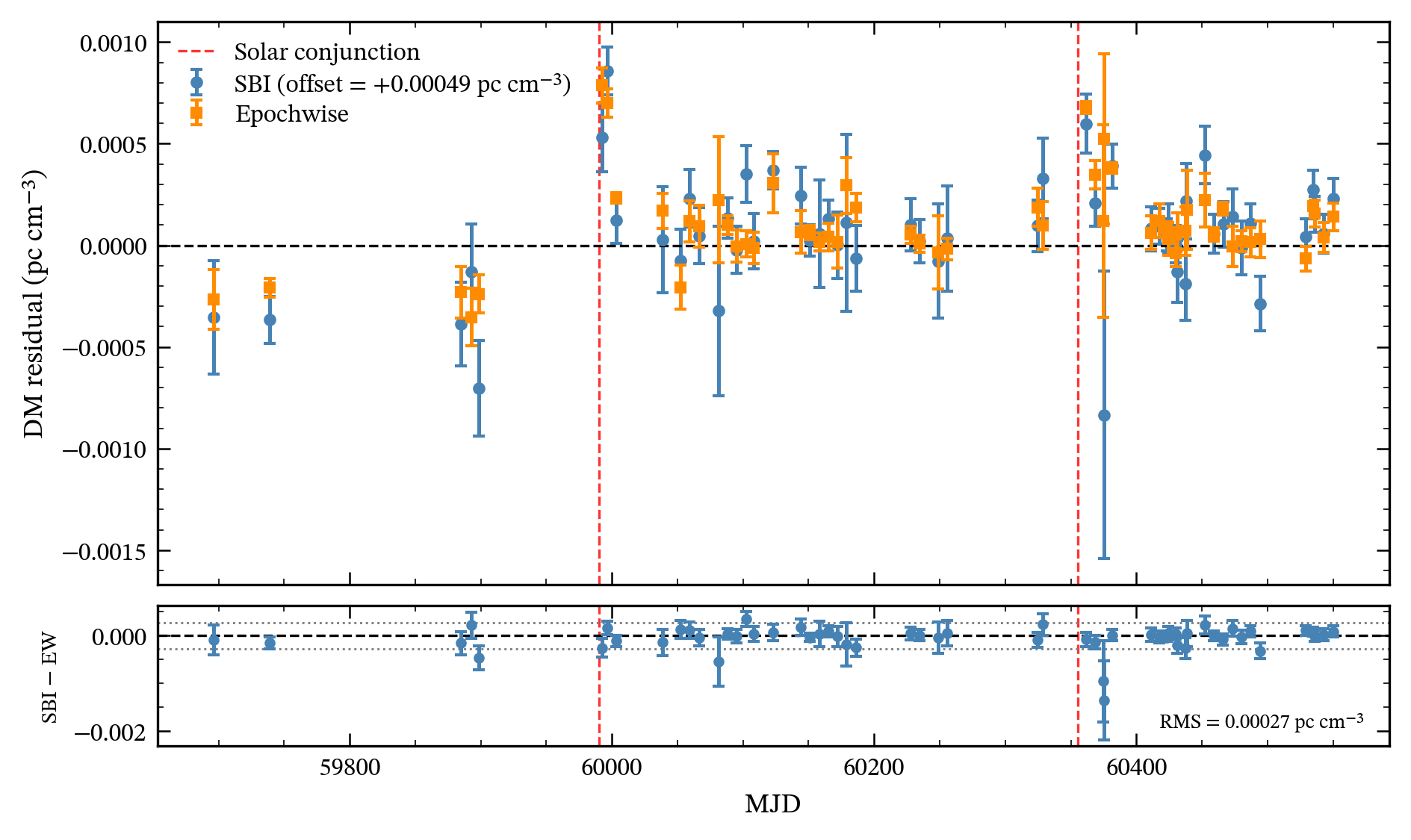}
\caption{SBI-inferred dispersion measures (blue) compared with the epoch-wise values (orange) from \citet{susarla2025} for PSR~J2145$-$0750, after removing the constant offset between the two absolute reference DMs. Both series show comparable enhancements near two epochs $\approx365$~days apart, consistent with successive solar conjunctions. \textit{Bottom:} epoch-matched difference, RMS $=2.7\times10^{-4}$~pc~cm$^{-3}$, reduced $\chi^2=0.74\pm0.19$ against the combined per-epoch uncertainties. The red lines indicate the epochs where solar conjunctions happened.}
\label{fig:dmcomp}
\end{figure*}

\subsection{Comparison with conventional epoch-wise analysis}
Dispersion measures are the most direct quantity to compare between the two methods, as they are physical and do not depend on the analysis convention. We compare our SBI DMs against the Epoch-wise values from the I-LOFAR ecliptic pulsar data release \citep{susarla2025, susarla_2025_dr} for the same number of epochs, both referenced to their own fixed template DM. We report the network's inferred DM residual directly, relative to the fixed template DM used throughout training. The conventional epochwise series of \citet{susarla2025} uses a different fixed reference, and we subtract the resulting constant offset before any comparison. The archive processing happens at topocentric channel frequencies, while the ISM disperses the signal at frequencies in the solar-system barycentric frame leaving an annually periodic kinematic bias in the measured DM. To first order, DM transforms as a single Doppler factor \citep{Pennuccietal2014,kulkarni2020}. An observation taken while Earth moves towards the pulsar at radial velocity $v_r$ yields an apparent DM increased by $\mathrm{DM}\times v_r/c$. We compute $v_r(t)$ from the I-LOFAR site and the pulsar's position \citep{astropy:2018} and subtract $\mathrm{DM}_\mathrm{ref}\times v_r(t)/c$ from each epoch from the SBI DMs. This produces a constant offset of 4.9$\times$10$^{-4}$ pc cm$^{-3}$ that we subtract and compare the variations directly.

\begin{figure}[!h]
    \centering
    \includegraphics[width=\linewidth]{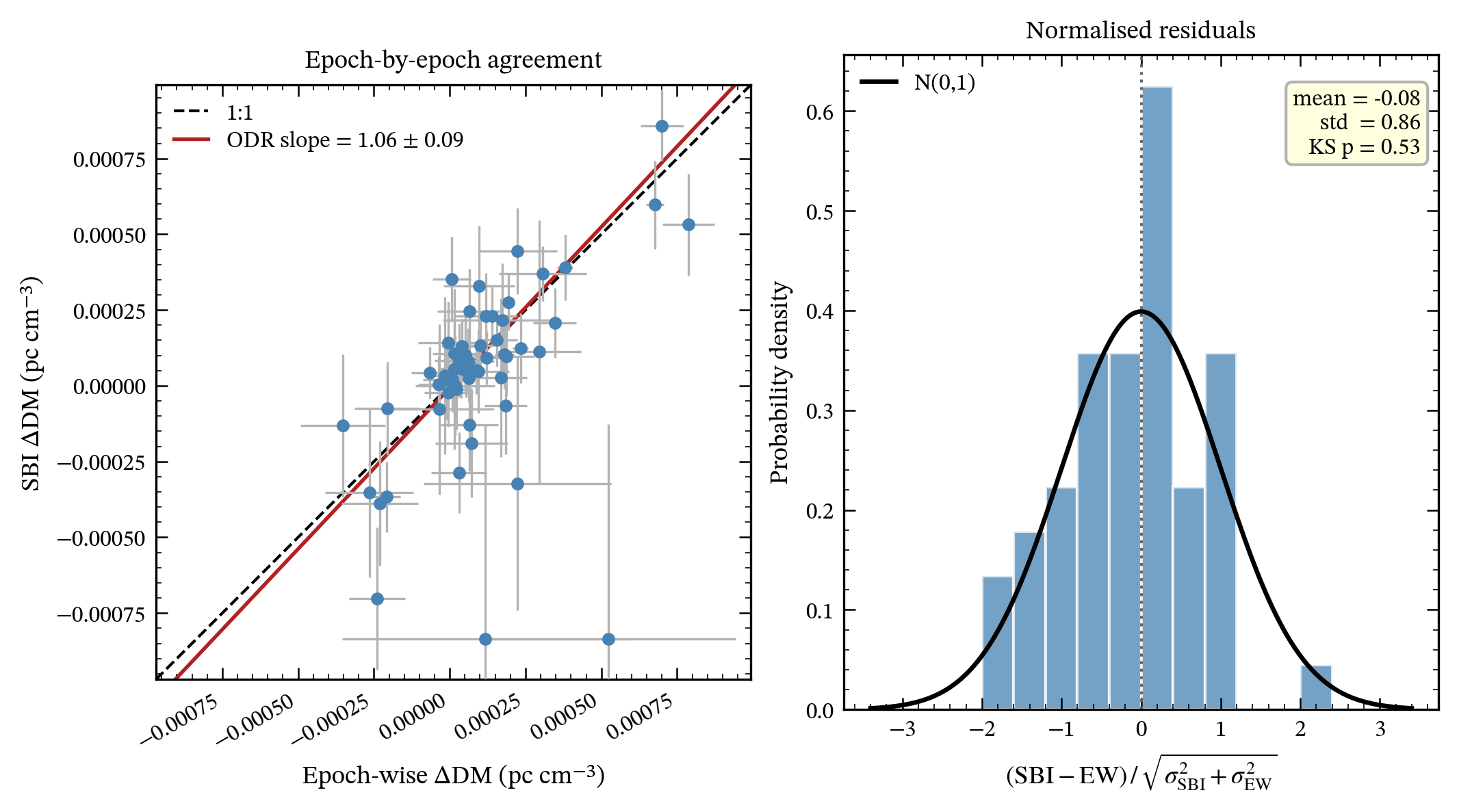}
    \caption{DM consistency checks between both analyses. The 1st panel shows how epoch-by-epoch agreement where the black dashed line shows a perfect agreement and the red solid line shows the ODR slope consistent with unity. The second panel shows the normalised residuals between both the time-series being consistent with standard normal distribution with a KS test p-value of 0.53 and SW p-value of 0.625. }
    \label{fig:dmconsistency}
\end{figure}

Visually, the two time-series in Figure~\ref{fig:dmcomp} show broad agreement. Neither shows a long term trend, and both show comparable DM enhancements near MJD~59990--60000 and MJD~60325--60375, $\approx365$~days apart and coincident with successive annual solar conjunctions \citep{Tiburzi2021,susarla2024}. The epoch-matched difference between the two series has an RMS of 2.7$\times$10$^{-4}$~pc~cm$^{-3}$ (shown in the bottom panel) with no significant trend across the dataset, consistent with the combined per-epoch uncertainties of the two methods. The two highest error-bar epochs near MJD~60375 are also the two worst ones in the conventional DM timeseries as well. These two have the lowest S/N in our dataset, with both of them hugging the prior boundary of S/N. Removing these biased epochs reduces the RMS to 1.6$\times$10$^{-4}$~pc~cm$^{-3}$.
To quantify the level of agreement, we look at the normalised residual DM in Fig.~\ref{fig:dmconsistency}: the difference in median predictions per epoch normalised by their combined uncertainties; the distribution of the normalised residual DM has a mean $-0.08 \pm 0.13$ and standard deviation $0.86 \pm 0.09$, statistically indistinguishable from a standard normal (Kolmogorov--Smirnov $p = 0.53$; Shapiro--Wilk $p = 0.63$), with a reduced $\chi^2$ of $0.74 \pm 0.19$. An errors-in-variables regression of the SBI values against the conventional ones gives a slope of $1.06 \pm 0.09$, consistent with unity at $0.7\sigma$. The measured standard deviation of $0.86$ matches that found for $\Delta$DM on simulated data (Sect.~\ref{sec:simul_valid}), indicating that the mild conservatism of the DM posteriors carries over from simulation to real observations.

Looking at the size of the uncertainties, we do find that the SBI DM uncertainties are on average $1.75\times$ larger than the conventional formal errors, reflecting marginalisation over the component widths, relative amplitudes and per-channel bandpass gains that template matching treats as exactly known. For PSR~J2145$-$0750, whose profile shows no reported change events, this additional conservatism is not accompanied by any reduction in bias. We note that both error estimates are consistent with the observed scatter between the two series (reduced $\chi^2 = 0.74\pm0.19$). This trade-off would be expected to favour marginalisation for pulsars whose profile stability cannot be assumed.

\section{Discussions and Conclusions}
\label{sec:conclusions}
We have presented a simulation-based inference approach to profile-domain pulsar timing, and applied it to I-LOFAR observations of PSR~J2145$-$0750. The method learns the frequency--phase portrait of the pulsar directly from simulations, and infers the dispersion measure, arrival time, and profile-evolution parameters for each observation without per-observation sampling. On real data, it recovers DMs consistent with a conventional analysis, including the solar-wind signal, and does so in under a second per observation after a single training stage.
The main advantage of this approach is that its cost is amortised. The shapelet-based Bayesian method introduced by \citet{lentati2017} performs one joint fit across all observations rather than a per-epoch measurement, with cost scaling sharply with dataset size ($>10^3$ parameters, requiring Hamiltonian Monte Carlo rather than nested sampling). This analysis must be repeated with addition of new observations, and this cost grows with the size of the dataset. Our method pays the cost once, in training, after which inference on any number of observations is negligible. This is well suited to the data volumes of the SKA era \citep{braun_ska}, where per-source sampling over large datasets becomes impractical, and where the profile evolution that this method models will be present across wide bands and at low frequencies.
Because the inference works on the full two-dimensional portrait rather than on frequency-collapsed arrival times, it also constrains the profile-evolution parameters directly. Existing wideband methods also parameterise frequency evolution \citep{Pennucci2019}, but do so in a separate template-construction stage whose result is then held fixed when measuring ToAs. The profile-evolution parameters are consequently not re-estimated per epoch, and their uncertainty does not propagate into the reported timing precision. For PSR~J2145$-$0750 we do not draw an astrophysical result from them, but the ability to measure them routinely opens a direction that conventional timing cannot follow. We note that the recovered spectral indices vary between epochs by more than their formal uncertainties, but intrinsic spectral variation is degenerate with epoch-dependent bandpass and propagation effects in a single reduced archive, and separating them requires calibration information beyond the data used in this work.

The present model has a deliberate scope. It captures the frequency evolution of the profile only; it is epoch-agnostic, and includes no time-domain timing model. The small structure remaining in the arrival-time residuals reflects time-domain effects that lie outside this scope. The profile is modelled with two components, and more complex profiles may require additional parameters. The posteriors are calibrated to within a few per cent, mostly in the conservative direction. None of these limitations affect the results presented here, and each points to a specific extension. To that end, we tested replacing the 2D CNN encoder with a channel-wise self-attention architecture (each frequency channel treated as a token, with an explicit frequency tag), trained on 20 million simulations. This encoder achieved a lower validation loss than the CNN and passed all posterior-predictive checks, though real-data DM uncertainties remained moderately larger (2.6x vs. 1.7x the epoch-wise errors). Because it does not assume a fixed channel grid, this architecture generalizes to models trained jointly across multiple telescopes with differing frequency coverage. We view this as a promising direction for future multi-frequency and multi-instrument SBI pipelines.

The natural next step is to incorporate a time-domain timing model into the same framework, so that spin and secular effects are inferred jointly with the frequency evolution. The method also extends directly to other pulsars, provided their frequency evolution can be simulated, and to other frequency bands including the low frequencies of SKA. More broadly, the same amortised, portrait-level approach can be brought to the binary pulsar search problem, where the computational cost of conventional methods is a growing limitation.
Simulation-based inference makes rigorous profile-domain timing fast enough for routine use. As pulsar datasets continue to grow, we expect amortised methods of this kind to become an increasingly practical part of the timing toolkit.

\section*{Acknowledgements}

SCS and GA would like to thank Dr Vanessa Graber (Royal Holloway, University of London) for providing access to GPU servers funded through a UKRI Future Leaders Fellowship (grant number MR/Y018257/1).
We would also like to acknowledge the support of the INTIME project (Grant No: UKRI1178). JPWV acknowledges support through NSF AccelNet award No.~2114721 and through the NANOGrav Physics Frontier Center through NSF award number 2607948. We used Claude (Anthropic) to assist with code review and manuscript editing.

\section*{Data Availability}

The I-LOFAR data that was used in this work is already public and is accessible at the zenodo archive \citep{susarla_2025_dr}. The trained posterior, the simulated validation sets, and the inferred parameter timeseries are deposited at \href{https://doi.org/10.5281/zenodo.21919057}{10.5281/zenodo.21919057} \citep{SBI_pulsars_susarla_zenodo}.
\section*{Code availability}
The simulator, embedding network, training pipeline and analysis scripts developed for this work are deposited at \href{https://doi.org/10.5281/zenodo.21919057}{10.5281/zenodo.21919057} \citep{SBI_pulsars_susarla_zenodo}. The analysis made use of \textsc{sbi} \citep{tejero2020}, \textsc{Pytorch} \citep{Paszke2019}, \textsc{PSRCHIVE} \citep{hotanpsrchive,straatenpsrchive}, \textsc{astropy} \citep{astropy:2018}, \textsc{numpy} \citep{Harrisetal2020}, \textsc{scipy} \citep{2020SciPy-NMeth} and \textsc{matplotlib} \citep{Matplotlib2007}.



\bibliographystyle{mnras}
\bibliography{freq_bib} 




\appendix

\section{DM consistency}
\label{sec:dmconsistency}
As the two analyses reference their DMs to different fixed template values, the series differ by an unknown constant offset $c$. Considering the SBI value at epoch $i$ as $\Delta DM^{SBI}_i \pm \sigma_{SBI,i}$ and the epoch-wise value as $\Delta DM^{EW}_i \pm \sigma_{EW,i}$, the maximum likelihood estimate of $c$ under Gaussian errors minimises $Q(c)$ which is defined as:
\begin{equation}
    Q(c) = \sum_i \frac{(\Delta DM^{SBI}_i + c - \Delta DM^{EW}_i)^2}{\sigma_{SBI,i}^2 + \sigma_{EW,i}^2}.
\end{equation}
Now, setting $\mathrm{d}Q/\mathrm{d}c = 0$ gives the inverse-variance weighted mean of the per-epoch differences,
\begin{equation}
    \hat{c} = \frac{\sum_i w_i (\Delta DM^{EW}_i - \Delta DM^{SBI}_i)}{\sum_i w_i},
    \qquad w_i = \frac{1}{\sigma_{SBI,i}^2 + \sigma_{EW,i}^2},
    \qquad \sigma_{\hat c} = \Bigl(\textstyle\sum_i w_i\Bigr)^{-1/2}.
\end{equation}
We obtain $\hat{c} = (4.9 \pm 0.2)\times10^{-4}$~pc~cm$^{-3}$ from all the matched epoch pairs. Given $\hat{c}$, we form the normalised residual at each matched epoch,
\begin{equation}
    z_i = \frac{\Delta DM^{SBI}_i + \hat{c} - \Delta DM^{EW}_i}{\sqrt{\sigma_{SBI,i}^2 + \sigma_{EW,i}^2}}.
\end{equation}
If the two analyses measure the same quantity and both sets of uncertainties are correctly sized, $\{z_i\}$ is drawn from a standard normal distribution ($\mathcal{N}(0,1)$). We test the $z_i$ against $\mathcal{N}(0,1)$ with both a Kolmogorov--Smirnov (KS) test and a Shapiro--Wilk (SW) test, and report the mean, standard deviation and reduced $\chi^2$ $= \langle z_i^2 \rangle$. The normalised residuals establish that the two series are consistent on average. To test whether they agree in \emph{scale}, we additionally regress $\Delta DM^{SBI}_i$ on $\Delta DM^{EW}_i$. Because both variables carry uncertainties, ordinary least squares would bias the slope towards zero; we therefore use orthogonal distance regression (ODR), which accounts for errors on both axes. A slope consistent with unity indicates that the two analyses track the same variations proportionally.


\bsp	
\label{lastpage}
\end{document}